\documentclass[conference,a4paper]{IEEEtran}
\IEEEoverridecommandlockouts
\usepackage{cite}
\usepackage{amsmath,amssymb,amsfonts}
\usepackage{algorithmic}
\usepackage{graphicx}
\usepackage{textcomp}
\usepackage{xcolor}
\usepackage[colorlinks]{hyperref}

\def\BibTeX{{\rm B\kern-.05em{\sc i\kern-.025em b}\kern-.08em
    T\kern-.1667em\lower.7ex\hbox{E}\kern-.125emX}}
\begin{document}

\title{SaRNet: A Dataset for Deep Learning Assisted Search and Rescue with Satellite Imagery}

\author{\IEEEauthorblockN{Michael Thoreau}
\IEEEauthorblockA{\textit{Department of Electrical and Computer Engineering} \\
\textit{New York University}\\
mjt9978@nyu.edu
}
\and
\IEEEauthorblockN{Frazer Wilson}
\IEEEauthorblockA{\textit{No Affiliation}\\
frazerwilson@mac.com}
}

\maketitle


\begin{abstract}
Access to high resolution satellite imagery has dramatically increased in recent years as several new constellations have entered service. High revisit frequencies as well as improved resolution has widened the use cases of satellite imagery to areas such as humanitarian relief and even Search and Rescue (SaR). We propose a novel remote sensing object detection dataset for deep learning assisted SaR. This dataset contains only small objects that have been identified as potential targets as part of a live SaR response. We evaluate the application of popular object detection models to this dataset as a baseline to inform further research. We also propose a novel object detection metric, specifically designed to be used in a deep learning assisted SaR setting.
\end{abstract}


\section{Introduction}


In-domain datasets are currently indispensible for applying machine learning models to real-world problems. One such problem is the search for missing persons in remote locations, where access can be difficult and timing is critical. So far, datasets for visual search and rescue (SaR) have mostly contained images taken by UAVs or light aircraft. This data cannot be simply translated to a satellite imagery setting, where there are currently no SaR datasets that we know of, due to greater diversity in viewpoints and relatively large target sizes. Modern high-resolution satellite constellations that can be tasked with imaging almost anywhere on the planet in a small number of hours, might soon enable a powerful complement to aerial searches, particularly in combination with recent advances in deep learning.

\begin{figure}[ht]
\begin{center}
  \includegraphics[width=0.8\linewidth]{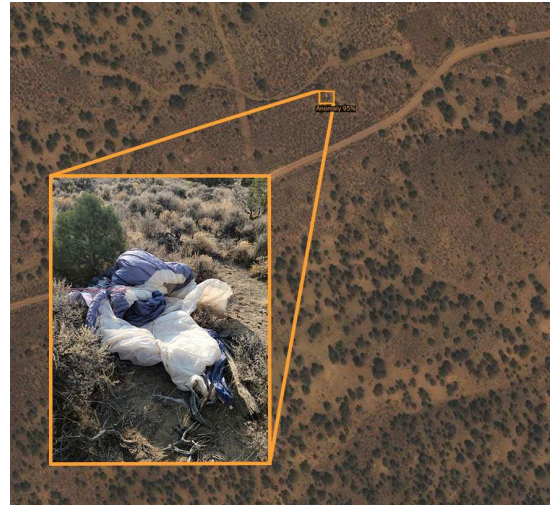}
\end{center}
  \caption{Inset - The paraglider wing as it was found. Main - The wing as detected by our prototype system.}
\label{fig:hero}

\end{figure}

We propose a novel object detection dataset, collected in a live search setting, that we use to demonstrate the concept of deep learning assisted SaR. The dataset was created during the search for a missing paraglider pilot, lost in a remote and mountainous area of the western United States. Over 500 volunteers labelled potential targets in high-resolution images using axis-aligned bounding boxes. The true target, as seen inset in figure \ref{fig:hero}, was found after a three week search and the labels generated for potential targets were saved. These images and annotations were post-processed to form this dataset of 2552 images. \\

Search and rescue via satellite imagery is a challenging application for off-the-shelf deep learning methods. Metrics typically used to evaluate a models performance on datasets such as MS-COCO \cite{DBLP:journals/corr/LinMBHPRDZ14} are very informative when the ground truth is fairly irrefutable and labelling is consistent, but have some issues when labels are noisy. Systems that use human verification as part of target acquisition can also generally tolerate a lower precision. We propose a new metric that is better suited to deep learning assisted SaR, that provides an intuitive way to choose a detection threshold for a given set of test images. We will evaluate a number of popular object detection models using this new metric on our dataset. Our contributions are as follows:

\begin{itemize}
    \item We present a novel dataset for satellite imagery based SaR, as well as;
    \item A novel and specially informative metric for object detection in a SaR setting, and;
    \item We perform an comparative study of popular object detection models trained and tested on this dataset.
\end{itemize}


\begin{figure}[ht]
    \centering
    \includegraphics[width=0.37\textwidth]{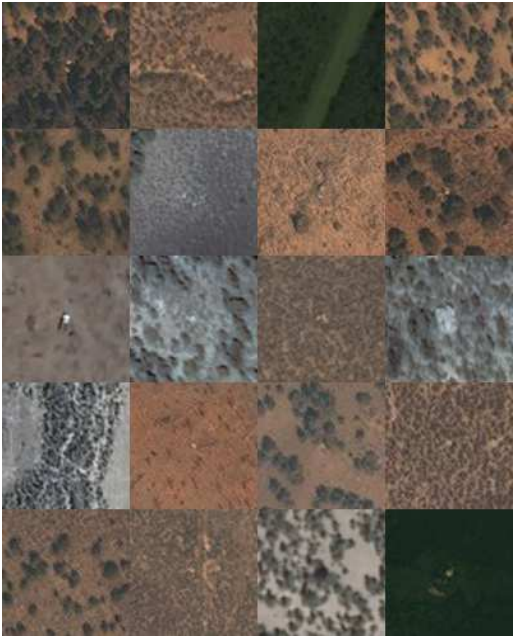}
    \caption{Representative training samples, selected at random.}
    \label{fig: training_samples}
\end{figure}

\section{Dataset Details}
\begin{figure}[ht]
    \centering
    \includegraphics[width=0.4\textwidth]{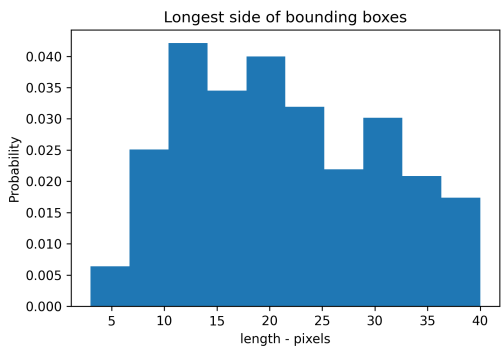}
    \caption{Distribution of the longest side of bounding boxes in the training set.}
    \label{fig: box_dist}
\end{figure}

This dataset contains 2552 images with a total of 4206 axis aligned bounding boxes of a single `target' class. Volunteers were instructed to label anything they think could be the missing paraglider wing and were provided with examples of similar objects visible in the source data. The wing is shown inset in figure \ref{fig:hero} as it was found after a three week search. A total of approximately 5000 annotations were originally generated, however bounding boxes with a height or width greater than 20 meters were discarded, along with the corresponding images. A mosaic of 20 labelled targets, selected at random, can be seen in figure \ref{fig: training_samples}. Each 1000x1000 image in the dataset is a jpeg rendered at a high quality factor corresponding to a 500m x 500m tile of satellite data, with a pixel pitch of 0.5m. The images have had all geographic and vendor information removed. The scale of the targets can be as small as 3-4 pixels in some cases, which is a particular challenge for machine learning models, as we will discuss in a later section. The bounding box labels in this dataset are generally slightly oversized relative to the target outlines. The distribution of the length of the longest side of the boxes in the training set can be seen in figure \ref{fig: box_dist}.


In contrast to most object detection datasets, including those in the remote sensing domain, the targets in this dataset cannot be considered a strong ground truth, however we will argue that they can still be used to train a useful object detector. Objects that might have been considered by labelers to be a potential target may have varied over time and between volunteers. This noise in the labels can be seen qualitatively in figure \ref{fig: training_samples}, where there is some variance in the obviousness of the targets.  For a dataset with a weaker ground truth, metrics based directly on false positives and false negatives are not as informative, as we will discuss in section \ref{sec: metrics}.



Examples have been split in to 70\%, 20\%, 10\% divisions for training, validation, and testing respectively, with annotations following the MS-COCO \cite{DBLP:journals/corr/LinMBHPRDZ14} convention. This dataset will be available online with example code: \url{https://github.com/michaelthoreau/SearchAndRescueNet}.

\section{Deep Learning Assisted SaR}

It is important to consider how the outputs of any model might be used in a wider system. Let us consider the case of the search \cite{binding_2020} that created this dataset, where targets were identified in a two step process. First, volunteers labeled potential targets in image tiles without contextual information. These labels were then verified by other volunteers with additional information such as historical satellite imagery for comparison. In this environment, a high number of false positives from the first step could be tolerated, as they would be filtered in the second step, with only a small number of potential targets being forwarded on to ground/air search teams. Anecdotally, we found that verification of the potential targets (second stage) was less tiring for volunteers than actively searching for targets in image tiles (first stage).

We propose that an object detection model could be used to replace or assist humans in the first stage of this SaR pipeline. In this proposed system, humans would initially label data in a new target domain, and a model would be trained to provide detections as search areas expanded. These detections could be used either as proposals in the first stage, highlighting potential targets and reducing strain on volunteers, or directly as potential targets for verification in the second stage. In both cases, search areas could be covered faster while keeping a human in the loop and reducing strain on resources.

\section{Prior Works}

A primary challenge for designing, evaluating, and deploying models to assist in SaR missions is the lack of applicable datasets. One popular dataset \cite{DBLP:journals/corr/abs-1711-10398} for remote sensing object detection has bounding box annotations for a variety of object categories, with the smallest being `small vehicle'. However this dataset contains aerial imagery with a finer pixel pitch than the proposed imagery. Fewer datasets exist for satellite based remote sensing data, and none currently available contain objects as small as those in the proposed dataset.

Detection of objects in remote imagery is a fairly well studied field, with state of the art methods \cite{7827088} using region based Convolutional Neural Networks (CNNs) to detect objects such as aircraft and oil tanks with a high degree of accuracy. These objects however have a scale in the order of 100s of pixels across, making the features learnt on common pre-training tasks such as ImageNet \cite{deng2009imagenet} very applicable and making transfer learning possible. One of the closest approaches \cite{8868719} to our problem achieves robust detection of small objects with highly variable backgrounds, however this domain has a much more variable object scale as well as significant viewpoint changes.

The use of deep learning to assist in SaR operations has been explored a number of times \cite{DBLP:journals/corr/abs-1904-11619,9311602, visual-based} but not as far as we can tell with satellite imagery. The authors of \cite{DBLP:journals/corr/abs-1904-11619} discuss how the human eye has incredible power to use context to discern true from false targets but is slow to scan images and can quickly become fatigued. This group \cite{DBLP:journals/corr/abs-1904-11619} also describes how small targets can be detected faster than by the human eye in some circumstances. We propose that an object detector can assist in SaR as a first step in a machine-human process, with humans verifying potential targets.


\section{Object Detection Metrics} \label{sec: metrics}
A key metric used in object detection is mean Average Precision (mAP) which is calculated as the area under the curve when precision is plotted against recall for all classes. Informed by the precision vs recall curve, practitioners applying object detection models choose a threshold that corresponds to a point on the precision recall curve that they deem most reasonable. Due to the nature of our dataset that does not have a strong ground truth, precision is not directly informative of the performance of the model on the task. In the appendix, figure \ref{fig: mosaic_fp} shows some `false positives' that degrade the apparent performance of the model in standard metrics but appear to be reasonable detections. We also found that when applying object detection in an SaR setting, we were setting the detection threshold based on the perceived density of detections. We propose a metric that can be directly related to the time cost of verifying candidate detections. We will define the detection density as the number of detections per square km. For a set of detections with confidence values $\mathcal{Y} = \{ y_1, y_2, ..., y_M\}$ estimated for a given image $x_i$ in a set $\mathcal{X} = \{x_1, x_2, ... , x_N  \}$ the detection density $D(\tau)$ can be found as:

\begin{equation}
    D(\tau) = \sum_{i = 1}^N \frac{|y_i > \tau|}{\textit{Area(}x_i\textit{)}} 
\end{equation}

Where $|y_i >\tau|$ denotes the number of detections corresponding to a given image with a confidence greater than $\tau$.  

Similar to mAP, we propose a metric that considers the average performance over a range of thresholds. In this case we will consider the recall of the detector at various detection densities between 0 and some reasonable number per square km, which can be chosen based on the dataset. We found that 20 detections per square km was a reasonable maximum density in this case. Recall was included in the metric as we found it remained fairly relevant in the presence of label noise, likely due to the labels being noisy but generally conservative. We call this metric the Average Recall-Density to 20 or AR-d20. The metric can be calculated as:

\begin{equation}
    \text{AR-d20} = \frac{1}{20}\sum_{k=0}^{20} \textit{Recall(}\tau_k\textit{)}
\end{equation}

Where $\tau_k$ is the lowest threshold that corresponds to a density lower than $k$ detections per square km and $\textit{Recall(}\tau_k\textit{)}$ is the recall for a given threshold. Crucially, this metric may inform the use of a particular model based on the human resources available for considering detections. In practice, the recall and detection densities can be estimated on a small amount of labeled test data in the target domain.

\section{Baseline Results}
To acquire a baseline solution, we fine tune a number of popular object detection models on the training data and evaluate the performance of each model on the test set using our new metric. The framework used to evaluate these models is Faster R-CNN\cite{DBLP:journals/corr/RenHG015} in which we consider three different feature extractors available in the Detectron2 \cite{wu2019detectron2} `model zoo'. As the targets are very small and the bounding box labels are relatively oversized, we relax the Intersection over Union (IoU) threshold for detections to $0.1$ to avoid the elimination of fair detections. We train all three models with a learning rate of 0.0001 for 5000 iterations. The optimiser was Adam, and each minibatch contained 4 images. All models were pre-trained on MS-COCO. We evaluate the baseline solution on the AR-d20 metric and present the results in table \ref{tab:ARd-20}.

\begin{figure}[h]
    \centering
    \includegraphics[width=0.49\textwidth]{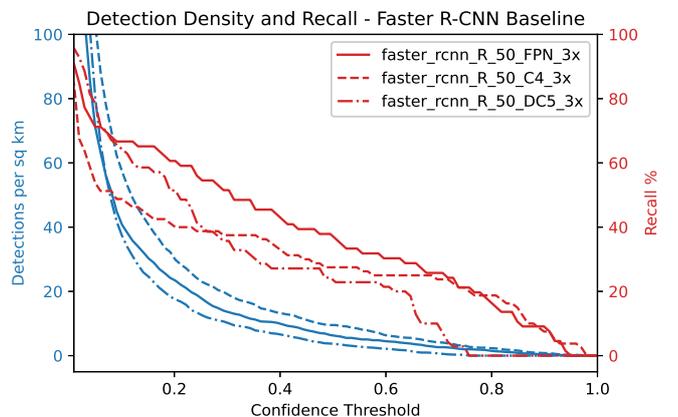}
    \caption{Density of detections and recall of 3 different Faster R-CNN models fine-tuned on this dataset.}
    \label{fig: density}
\end{figure}

The highest scoring model `R\_50\_FPN' includes a 50 layer resnet \cite{he2015deep} feature extractor. This model, which includes a Feature Pyramid Network \cite{lin2017feature}, had an average recall of 41.8\% for detection densities between 0 and 20 detections per square km. Figure \ref{fig: density} shows detection density and recall plotted for all three models at a range of confidence thresholds. The FPN model has a significantly higher recall than the other models but a similar detection density across most thresholds, making it a compelling choice. Qualitative detection results from applying this model to the validation set can be seen in figure \ref{fig: dets} in the appendix.

\begin{table}[ht]
    \centering
    \begin{tabular}{c|c}
        model & ARd-20 \\
        \hline 
        faster\_rcnn\_R\_50\_FPN\_3x & \textbf{41.82} \\
        faster\_rcnn\_R\_50\_C4\_3x & 28.88 \\
        faster\_rcnn\_R\_50\_DC5\_3x & 35.46\\
    \end{tabular}
    \caption{Average Recall-density scores for three popular models.}
    \label{tab:ARd-20}
\end{table}

\section{Conclusion}

We presented a novel dataset that we hope will inform future research into satellite imagery based SaR. We introduced a new metric that may assist in the application of object detectors to SaR problems and presented a baseline model trained on this dataset to demonstrate the deep learning assisted SaR concept. We believe that satellite based SaR is an emerging field that may someday be used to save lives and bring closure to families of missing persons. We firmly believe in the applicability of this technology to the search for missing aircraft, watercraft, and a variety of other targets we cannot yet imagine.


\section*{Acknowledgment}

The authors would like to thank Planet Labs, Airbus Defence and Space, and Maxar Technologies for providing satellite imagery. The authors would also like to thank the 500+ volunteers who labelled the data, as well as all those who contributed to the physical air and land search. \\



\appendices
\section{Supplementary figures}

\begin{figure}[ht]
    \centering
    \includegraphics[width=0.46\textwidth]{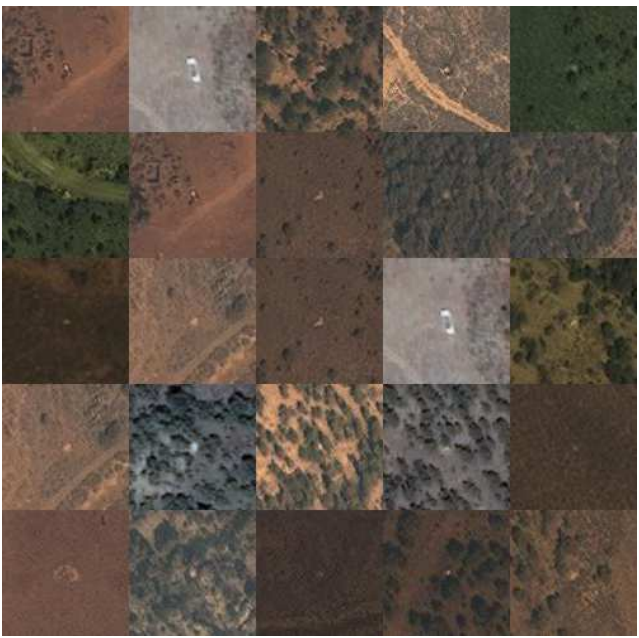}
    \caption{A selection of `False Positives' on the test set. These are objects that were detected by the model but missed by volunteers.}
    \label{fig: mosaic_fp}
\end{figure}

{\small
\bibliographystyle{ieee}
\bibliography{egbib}

\begin{thebibliography}{10}\itemsep=-1pt

\bibitem{binding_2020}
L.~Binding.
\newblock Body of well-known paraglider found in us mountains - weeks after he
  went missing.
\newblock {\em Sky News}, Sep 2020.

\bibitem{9311602}
G.~Castellano, C.~Castiello, C.~Mencar, and G.~Vessio.
\newblock Preliminary evaluation of tinyyolo on a new dataset for
  search-and-rescue with drones.
\newblock In {\em 2020 7th International Conference on Soft Computing Machine
  Intelligence (ISCMI)}, pages 163--166, 2020.

\bibitem{deng2009imagenet}
J.~Deng, W.~Dong, R.~Socher, L.-J. Li, K.~Li, and L.~Fei-Fei.
\newblock Imagenet: A large-scale hierarchical image database.
\newblock In {\em 2009 IEEE conference on computer vision and pattern
  recognition}, pages 248--255. Ieee, 2009.

\bibitem{visual-based}
S.~Gotovac, D.~Zelenika, Z.~Marusic, and D.~Božić-Štulić.
\newblock Visual-based person detection for search-and-rescue with uas: Humans
  vs. machine learning algorithm.
\newblock {\em Remote Sensing}, 12, 10 2020.

\bibitem{he2015deep}
K.~He, X.~Zhang, S.~Ren, and J.~Sun.
\newblock Deep residual learning for image recognition, 2015.

\bibitem{DBLP:journals/corr/LinMBHPRDZ14}
T.~Lin, M.~Maire, S.~J. Belongie, L.~D. Bourdev, R.~B. Girshick, J.~Hays,
  P.~Perona, D.~Ramanan, P.~Doll{\'{a}}r, and C.~L. Zitnick.
\newblock Microsoft {COCO:} common objects in context.
\newblock {\em CoRR}, abs/1405.0312, 2014.

\bibitem{lin2017feature}
T.-Y. Lin, P.~Dollár, R.~Girshick, K.~He, B.~Hariharan, and S.~Belongie.
\newblock Feature pyramid networks for object detection, 2017.

\bibitem{7827088}
Y.~{Long}, Y.~{Gong}, Z.~{Xiao}, and Q.~{Liu}.
\newblock Accurate object localization in remote sensing images based on
  convolutional neural networks.
\newblock {\em IEEE Transactions on Geoscience and Remote Sensing},
  55(5):2486--2498, 2017.

\bibitem{DBLP:journals/corr/RenHG015}
S.~Ren, K.~He, R.~B. Girshick, and J.~Sun.
\newblock Faster {R-CNN:} towards real-time object detection with region
  proposal networks.
\newblock {\em CoRR}, abs/1506.01497, 2015.

\bibitem{8868719}
M.~Schembri and D.~Seychell.
\newblock Small object detection in highly variable backgrounds.
\newblock In {\em 2019 11th International Symposium on Image and Signal
  Processing and Analysis (ISPA)}, pages 32--37, 2019.

\bibitem{wu2019detectron2}
Y.~Wu, A.~Kirillov, F.~Massa, W.-Y. Lo, and R.~Girshick.
\newblock Detectron2.
\newblock \url{https://github.com/facebookresearch/detectron2}, 2019.

\bibitem{DBLP:journals/corr/abs-1711-10398}
G.~Xia, X.~Bai, J.~Ding, Z.~Zhu, S.~J. Belongie, J.~Luo, M.~Datcu, M.~Pelillo,
  and L.~Zhang.
\newblock {DOTA:} {A} large-scale dataset for object detection in aerial
  images.
\newblock {\em CoRR}, abs/1711.10398, 2017.

\bibitem{DBLP:journals/corr/abs-1904-11619}
K.~Yun, L.~Nguyen, T.~Nguyen, D.~Kim, S.~Eldin, A.~Huyen, T.~Lu, and E.~Chow.
\newblock Small target detection for search and rescue operations using
  distributed deep learning and synthetic data generation.
\newblock {\em CoRR}, abs/1904.11619, 2019.

\end{thebibliography}
}

\onecolumn
\begin{figure}[ht]
    \includegraphics[width=0.99\textwidth]{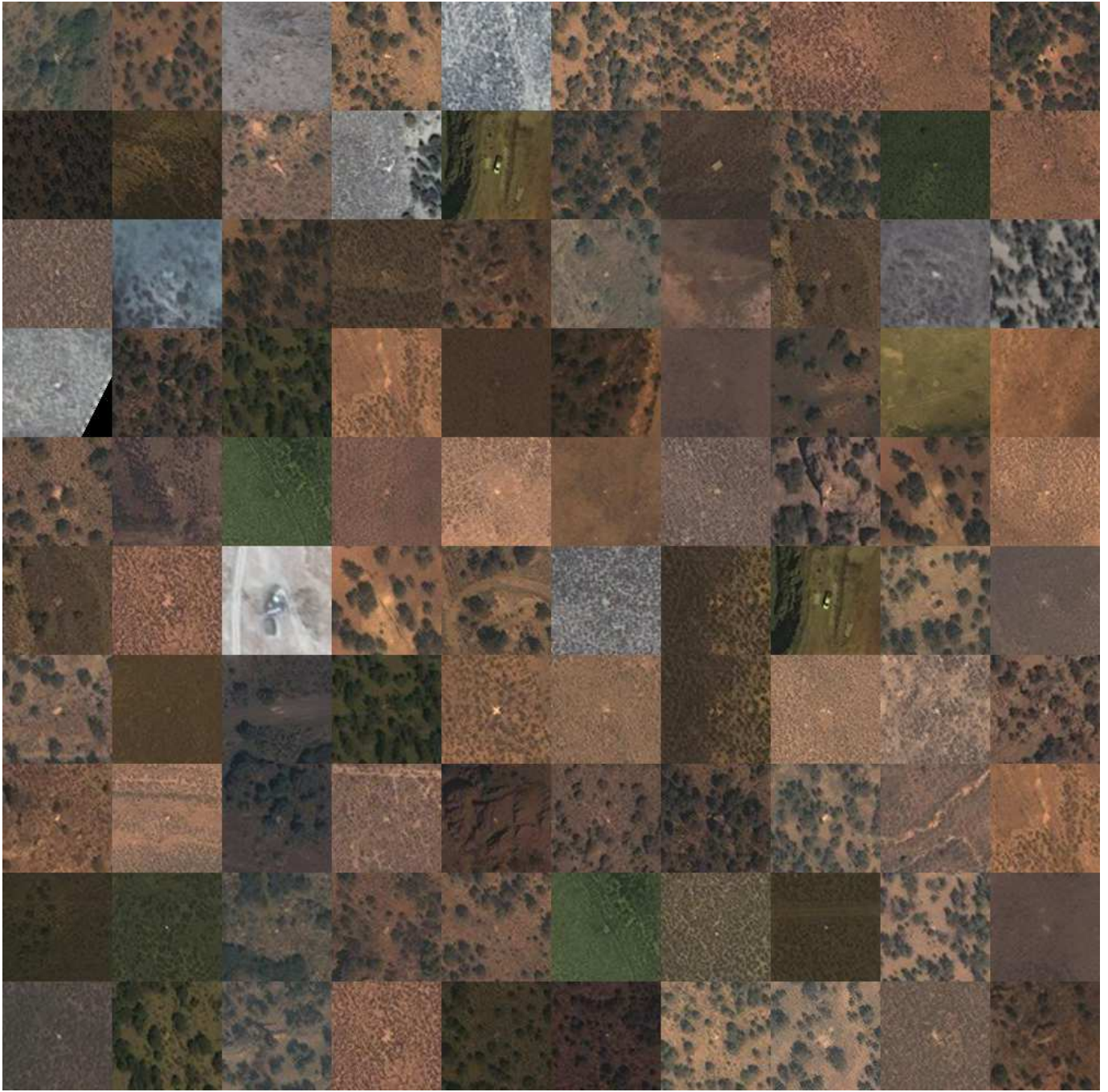}
    \caption{Example detections from the best performing model on the validation set. Each tile in the mosaic is a fixed size and is centred on the detection. Some examples are shown multiple times due to overlapping bounding boxes of very different sizes.}
    \label{fig: dets}
\end{figure}

\end{document}